\documentclass[a4paper, 11pt]{article}
\usepackage[top=1in, bottom=1in, left=1in, right=1in]{geometry}
\usepackage{graphicx}
\usepackage{amsthm}
\usepackage{amsmath}
\usepackage{amsfonts}
\usepackage{amssymb}
\usepackage{mathtools}
\usepackage{pdfpages}
\usepackage{natbib}
\usepackage{fancyhdr}
\usepackage[english]{babel}
\usepackage{txfonts}
\usepackage{gensymb} 
\usepackage{caption}
\usepackage{listings,relsize} 
\usepackage{setspace}
\usepackage{tcolorbox}
\usepackage{chngcntr}
\usepackage{lineno}
\usepackage{booktabs}
\usepackage{lineno}
\usepackage{ulem}
\usepackage{url}

\newcommand{\jfdel} {\bgroup\markoverwith{\textcolor{blue}{\rule[0.5ex]{2pt}{1.5pt}}}\ULon}

\begin{document}


\title{\Large{\textbf{Phylogenetic least squares estimation without genetic distances}}}

\author{\small{\textbf{Peter B. Chi}$^{1}$ and \textbf{Volodymyr M. Minin}$^{2,3}$}\\
\footnotesize{$^{1}$Department of Mathematics and Statistics, Villanova University, Villanova, PA, U.S.A.}\\
\footnotesize{$^{2}$Department of Statistics, University of California, Irvine, CA, U.S.A.}\\
\footnotesize{$^{3}$Center for Complex Biological Systems, University of California, Irvine, CA, U.S.A.}\\
}
\date{}

\maketitle

%

\label{firstpage}

\begin{abstract}
Least squares estimation of phylogenies is an established family of methods with good statistical properties.
State-of-the-art least squares phylogenetic estimation proceeds by first estimating a distance matrix, which is then used to determine the phylogeny by minimizing a  squared-error loss function. Here, we develop a method for least squares phylogenetic inference that does not rely on a pre-estimated distance matrix. Our approach allows us to circumvent the typical need to first estimate a distance matrix by forming a new loss function inspired by the phylogenetic likelihood score function; in this manner, inference is not based on a summary statistic of the sequence data, but directly on the sequence data itself. 
We use a Jukes-Cantor substitution model to show that our method leads to improvements over ordinary least squares phylogenetic inference, and is even observed to rival maximum likelihood estimation in terms of topology estimation efficiency.
Using a Kimura 2-parameter model, we show that our method also allows for estimation of the global transition/transversion ratio simultaneously with the phylogeny and its branch lengths. 
This is impossible to accomplish with any other distance-based method as far as we know.
Our developments pave the way for more optimal phylogenetic inference under the least squares framework, particularly in settings under which likelihood-based inference is infeasible, including when one desires to build a phylogeny based on information provided by only a subset of all possible nucleotide substitutions such as synonymous or non-synonymous substitutions. 
\end{abstract}


\section{Introduction}
\par Model-based methods for inferring phylogenies include distance-based methods and likelihood-based methods. While likelihood-based methods such as Bayesian and maximum likelihood approaches have highly attractive statistical behavior in general, they can suffer in terms of speed and computational burden. Conversely, distance-based methods are generally the easiest to program, are very fast, and are well-justified statistically, leading to their continued popularity \citep[Chapter 11]{Felsenstein_2004}. 

\par Additionally, in certain circumstances, distance-based methods may be the only viable option, such as when constructing a phylogeny based on a pairwise alignment \citep{Thorne_1992, Xia_2013, Xia_2017}, or when one is interested in building phylogenies based only on certain types of substitutions; for example, in an investigation of drug resistance by the HIV virus, \citet{Lemey_2005} constructed phylogenies of the HIV \textit{pol} gene based on synonymous distances (i.e. those that result in the same amino acid) and non-synonymous distances (i.e. those that change the amino acid) separately, using the Syn-SCAN algorithm \citep{Gonzales_2002}. In doing so, they found that the phylogeny based on synonymous distances matched the known transmission history due to contact tracing, whereas the phylogeny based on non-synonymous distances did not, therefore enabling an argument that selective pressures had acted on the HIV genome. This notion of labelled substitutions was generalized in \citep{Minin_2008, Obrien_2009}, and the framework established therein was used in a method for the detection of recombination that can simultaneously avoid signals of convergent evolution, by comparing trees built based on synonymous amino acid substitutions vs. non-synonymous amino acid substitutions \citep{Chi_2015}.

\par Therefore, the attractiveness of distance-based methods remains high to this day.  As noted by \citet[Chapter 11]{Felsenstein_2004}, least squares phylogenetic inference generally performs almost as well as likelihood-based methods in terms of bias and variance, and better than all other methods. Our aim in this work is to devise an improvement to least squares phylogenetic inference that will bring it closer to the optimality of likelihood-based inference while maintaining the flexibilities of distance-based methods.

\par Similar to estimation of a least squares regression line, least squares phylogenetic inference aims to minimize the squared differences between observed and predicted quantities, where the former in this setting are represented by the matrix of pairwise evolutionary distances between each pair of molecular sequences, and the latter is a matrix of pairwise tree distances based on the candidate tree, dictated by the tree topology and branch lengths \citep{Cavalli-Sforza_1967}. This makes least squares phylogenetic inference a two-stage procedure, since the distance matrix must first be estimated from the data, before tree estimation can then proceed based on this estimated distance matrix. In other words, inference is actually based on a summary statistic of the data (and one that is not necessarily a sufficient statistic), resulting in loss of information by the estimation procedure, and thus poorer performance \citep{Penny_1992}. 

\par Here, we aim to improve upon current least squares phylogenetic inference based on a modification of the least squares criterion, or loss function. Rather than treating distances as fixed quantities, our new loss function relies on conditional expectations that depend both on the data and the candidate tree. Thus, the actual sequence data are considered simultaneously with the candidate trees, as opposed to first summarizing the sequence data into the distance matrix. Although our new proposed distances still use only pairwise comparisons (as opposed to accounting for the correlation which may exist in higher-order comparisons), we show that our new loss function optimization results in better estimation of a phylogeny, including branch lengths, than using the usual least squares criterion, under a variety of simulation settings.


\par We begin with the JC69 model of nucleotide substitution \citep{Jukes_1969}, first examining branch length estimation over the correct tree, thus temporarily setting aside the usual issues of searching through tree space, and evaluating whether the use of our new loss function results in improved branch length estimation in terms of bias and variance. We then examine topology estimation and evaluate whether the use of our new loss function results in correctly identifying the topology more frequently than use of the original loss function, also including maximum likelihood estimation as a state-of-the-art benchmark. Finally, we also explore branch length estimation under one additional model of nucleotide substitution, the K80 model \citep{Kimura_1980}, to demonstrate the flexibility of our approach.

\section{Methods}
Consider a molecular sequence alignment, ${\boldsymbol Y}$, which is an $n \times L$ matrix composed of row vectors ${\boldsymbol y_1}, \ldots, {\boldsymbol y_n}$, where $n$ is the number of taxa/species, or genetic sequences of interest, and $L$ is the length, or number of sites in the sequence alignment. Then, ${\boldsymbol y_k} = (y_{k1}, \ldots, y_{kL})$, where each $y_{ks}$ for $s \in \{1, \ldots, L\}$ takes on a value determined by the type of data; for DNA nucleotide sequences, these values would be $\{A, C, G, T\}$. We assume each column of ${\bf Y}$ to be an independently evolving site, under an irreducible, reversible continuous-time Markov chain with infinitesimal generator ${\bf \Lambda} = \{\lambda_{ij}\}_{M \times M}$, and stationary distribution ${\boldsymbol\pi} = (\pi_1, \ldots, \pi_M)$, where $M$ is the size of the state space (in the case of DNA nucleotides, $M=4$). In the phylogenetic setting, the infinitesimal generator ${\bf \Lambda}$ generally consists of parameters specific to any given molecular substitution model; for example, in the JC69 model, the infinitesimal generator is defined by a single parameter $\alpha$, and we use the following parameterization:

\begin{align}
\label{jc69rate}
\boldsymbol{\Lambda}_{JC} &= \frac{1}{4} \begin{pmatrix}
-3 \alpha & \alpha & \alpha & \alpha \\
\alpha & -3 \alpha & \alpha & \alpha \\
\alpha & \alpha & -3 \alpha & \alpha \\
\alpha & \alpha & \alpha & -3 \alpha \\
\end{pmatrix}.
\end{align}
In general, we will refer to the vector of all existing parameters in the substitution model of interest as $\boldsymbol{\theta}$. We also define $\lambda_i = \sum_{j \neq i} \lambda_{ij}$ as the rate of leaving any given state $i$; for our parameterization of the JC69 model, $\lambda_i=(3/4)\alpha$ for all $i$.

Traditionally, the evolutionary distance between a pair of sequences is defined as the expected number of substitutions per site, and is typically calculated by first considering the finite-time transition probability matrix ${\bf P}(t) = \{p_{ij}(t)\}_{M \times M}$. Each $p_{ij}(t)$ in this matrix is the probability that any site in one sequence is in state $j$ given that it is in state $i$ in the other sequence, when the two sequences are separated by time $t$. The finite-time transition probability matrix is obtained by noting that $\boldsymbol{P}(t) = e^{\boldsymbol{\Lambda}t}$ \citep{Guttorp_1995}. For the JC69 model, this gives:

  \begin{align}
  \label{jc69prob}
\boldsymbol{P}_{JC}(t) &= \frac{1}{4} \begin{pmatrix}
1+3 e^{- \alpha t} & 1-e^{-\alpha t} & 1-e^{-\alpha t} & 1-e^{-\alpha t} \\
1-e^{-\alpha t} & 1+3 e^{- \alpha t} & 1-e^{-\alpha t} & 1-e^{-\alpha t} \\
1-e^{-\alpha t} & 1-e^{-\alpha t} & 1+3 e^{- \alpha t} & 1-e^{-\alpha t} \\
1-e^{-\alpha t} & 1-e^{-\alpha t} & 1-e^{-\alpha t} & 1+3 e^{- \alpha t}
\end{pmatrix}.
  \end{align}
Then, we note that the average number of state changes in any stationary continuous-time Markov chain is equal to:

\begin{align}
\sum_{i=1}^M \pi_i \lambda_i t,
\end{align}
where for the JC69 model, $\pi_i=1/4$, and $\lambda_i$ is equal to $(3/4)\alpha$ as mentioned above, for all $i$. Since the definition of an evolutionary distance is equivalent to the average number of state changes, this means that the JC69 evolutionary distance is equal to:

\begin{align}
d_{JC} = \frac{3}{4} \alpha t.
\end{align}
Finally, from the transition probability matrix, the probability of observing any change of state after time $t$ in the JC69 model is $p=3/4 - (3/4)e^{-\alpha t} = 3/4 - (3/4)e^{-(4/3)d_{JC}}$. Solving for $d_{JC}$ gives rise to a method of moments evolutionary distance estimator of: 

\begin{align}
\hat{d}_{JC} = -\frac{3}{4} log \left(1 - \frac{4}{3} \hat{p}\right),
\end{align}
where $\hat{p}$ is the empirical proportion of sites that differ between the pair of sequences in question. The evolutionary distances can thus be calculated for each pair of sequences in the molecular sequence alignment.



\begin{figure}[htb]
\begin{center}
\includegraphics[width=0.6\textwidth]{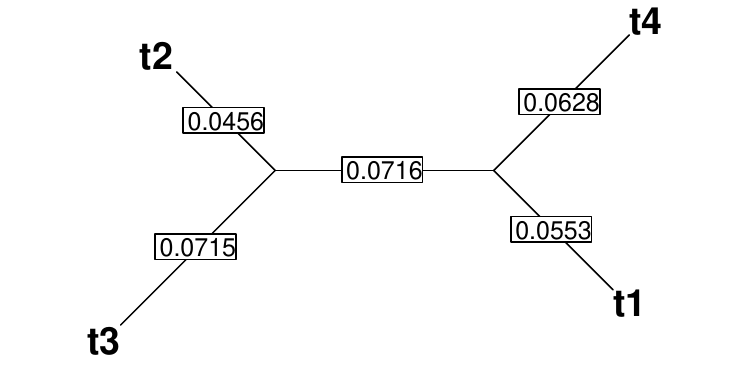}
\end{center}
\caption{\textbf{An unrooted, bifurcating phylogeny}. This particular topology indicates that taxa ``t1'' and ``t4'' are more closely related
to each other than to taxa ``t2'' and ``t3,'' and vice-versa. The numbers placed on each branch indicates
its length.}
\label{fig:tree4-1}
\end{figure}

Least squares phylogenetic inference then proceeds by finding the phylogenetic tree that best fits the pairwise evolutionary distances calculated from the molecular sequence data. A phylogenetic tree is an object that represents the evolutionary relationships between a set of species, or taxa; an example is shown in Figure \ref{fig:tree4-1}. This figure depicts an unrooted tree, in which the direction of evolutionary time is unknown. While methods for inferring rooted trees and for rooting unrooted trees exist, in this work we focus our attention to inferring unrooted trees. 

On a given phylogeny, we define the tree distance $t_{kl}({\boldsymbol\tau}, {\bf b})$ between taxa $k$ and $l$ as the sum of the branch lengths between them on the particular phylogeny determined by ${\boldsymbol{\tau}}$, the topology, and $\boldsymbol{b}$, the set of branch lengths on the topology. For example, tree distance between taxa t3 and t4 in Figure \ref{fig:tree4-1} is $0.0715 + 0.0716 + 0.0628 =0.2062$. Next, recall from above that we have estimated $\hat{d}_{kl}$ from our DNA sequence data. Then, the quantity

\begin{eqnarray}
\label{eqn:L1}
\mathcal{L}_1(\boldsymbol{d}, \boldsymbol{b}, \boldsymbol{\tau}) = \sum_{k>1} \left[\hat{d}_{kl} - t_{kl}(\boldsymbol{\tau}, \boldsymbol{b})\right]^2,
\end{eqnarray}
is the usual least squares criterion, adapted to the context of phylogenetics. 
The solution $(\hat{\boldsymbol{\tau}}, \hat{\boldsymbol{b}})$ to 
\begin{eqnarray}
{\underset{\boldsymbol{\tau}, \boldsymbol{b}} {\text{argmin }}} \mathcal{L}_1(\boldsymbol{d}, \boldsymbol{b}, \boldsymbol{\tau})
\end{eqnarray} 
then gives the ordinary least squares phylogeny.

\par In this work, we propose an improvement to least squares phylogenetic inference by considering a new loss function. Our new loss function bypasses the need for the intermediate step of estimating the distance matrix, and instead considers the sequence alignment directly. First, we define $N_{t_{kl}(\boldsymbol{\tau},\boldsymbol{b})}$ as the number of substitutions between taxa $k$ and $l$, along the tree defined by $\boldsymbol{\tau}$ and $\boldsymbol{b}$. Then, $\mathbb{E}(N_{t_{kl}(\boldsymbol{\tau},\boldsymbol{b})}|y_{ks}, y_{ls})$ is the expected number of substitutions between taxa $k$ and $l$, given their observed states at site $s$, and a chosen mutational model. The empirical average number of substitutions across the alignment is 
\begin{eqnarray}
\label{eqn:robustd}
e_{kl}(\boldsymbol{\tau, b, \theta}) = \frac{1}{L} \sum_{s=1}^L \mathbb{E}(N_{t_{kl}(\boldsymbol{\tau},\boldsymbol{b})}|y_{ks}, y_{ls}).
\end{eqnarray}
Finally, we define our new loss function
\begin{eqnarray}
\label{eqn:newloss}
\mathcal{L}_2(\boldsymbol{Y, b, \tau, \theta}) = \sum_{k>l} \bigg[e_{kl}(\boldsymbol{\tau, b, \theta})-t_{kl}(\boldsymbol{\tau},\boldsymbol{b})\bigg]^2,
\end{eqnarray}
which we note is similar to $\mathcal{L}_1$ in Equation (\ref{eqn:L1}), but with the replacement of the original model-based estimates of distance $\hat{d}_{kl}$ with the so-called robust distance $e_{kl}$ \citep{Minin_2008, Obrien_2009}. Explicitly, $e_{kl}$ depends on $(\boldsymbol{y_k, y_l, \theta, \tau, b})$ simultanously, whereas $\hat{d}_{kl}$ is a summary statistic of $(\boldsymbol{y_k,y_l})$ based on the previously estimated model parameters $\boldsymbol{\theta}$; we posit that optimization of our $\mathcal{L}_2$ loss function will result in better phylogenetic inference performance since the robust distances do not suffer from the same loss of information that traditional evolutionary distances do. Details for the calculation of (\ref{eqn:robustd}) in generality were demonstrated by \cite{Ball_2005}, \cite{Minin_2008} and \cite{Obrien_2009}. Additionally, we show analytic formulas for this quantity under the JC69 model in the Supplementary Materials. Under the parameterization of Equation (\ref{jc69rate}), we assume that $\alpha=(4/3)$, so that time is equal to the expected number of substitutions per site and thus our estimated robust distances can be interpreted directly as branch lengths in terms of evolutionary time. 

\par We further note that, by definition, $\mathbb{E}(N_{t_{kl}(\boldsymbol{\tau},\boldsymbol{b})}) = t_{kl}(\boldsymbol{\tau},\boldsymbol{b})$, because phylogenetic branch length is defined as the expected number of substitutions between two nodes. Next, recognizing that by the law of iterated expectations,
\begin{align}
\mathbb{E}[\mathbb{E}(N_{t_{kl}(\boldsymbol{\tau},\boldsymbol{b})} | y_{ks}, y_{ls})]
= \mathbb{E}(N_{t_{kl}(\boldsymbol{\tau},\boldsymbol{b})}),
\end{align}
then for any substitution model such that $\mathbb{E}(N_{t_{kl}(\boldsymbol{\tau},\boldsymbol{b})})$ exists and is finite, and with the assumption that each site of the sequence alignment evolves independently, by the Strong Law of Large Numbers we thus have, 
\begin{align}
\frac{1}{L} \sum_{s=1}^L \mathbb{E}(N_{t_{kl}(\boldsymbol{\tau},\boldsymbol{b})} | y_{ks}, y_{ls}) - \mathbb{E}(N_{t_{kl}(\boldsymbol{\tau},\boldsymbol{b})}) \xrightarrow{\text{a.s.}} 0
\label{loss_limit}
\end{align}
as $L \rightarrow \infty$. Therefore, under the correct substitution model, our least squares criterion $\mathcal{L}_2$ should be minimized at exactly 0 when we have identified the correct tree, as the sequence length $L \rightarrow \infty$. While we of course will never have the substitution model exactly correct, nor will we ever have infinite sequence length, this result suggests minimizing our new loss function will result in statistically consistent estimation. 
Moreover, \citet{Minin_2011} showed that for a pairwise alignment under the Jukes-Cantor model, setting the right-hand side of \eqref{loss_limit}
 to zero is equivalent to setting the derivative of the log-likelihood (score function) to zero. Therefore, in the special case of two sequences, minimizing our loss function \eqref{eqn:newloss} is equivalent to maximizing the log-likelihood of the Jukes-Cantor model.

\par Under the $\mathcal{L}_1$ loss function in Equation (\ref{eqn:L1}), the least squares solution can be obtained through standard matrix algebra approaches, as a phylogeny can be fully represented by the collection of all $t_{kl}(\boldsymbol{\tau},\boldsymbol{b})$, and the $\hat{d}_{kl}$ are constants, once they are obtained from the data \citep{Felsenstein_2004}. However, this is unfortunately not the case for our new $\mathcal{L}_2$ loss function in Equation (\ref{eqn:newloss}), since $e_{kl}$ depends on $\boldsymbol{\tau}$ and $\boldsymbol{b}$. Thus, iterative numeric approaches must be used. Here, we use the {\tt{nlminb}} routine \citep{Gay_1990, Fox_1997}, as this showed the best performance out of all routines that we explored (see Supplementary Material). For convenience, we used the {\tt{R}} wrapper package {\tt{optimx}} to call this routine \citep{Nash_2011, Nash_2014}.

\par Additionally, there exists a concern that $\mathcal{L}_2 \rightarrow 0$ as each $t_{kl}(\boldsymbol{\tau},\boldsymbol{b}) \rightarrow \infty$ (a proof of this in the case of the JC69 model is included in the Supplementary Materials). Thus, although the loss function would indeed be minimized at infinite branch lengths, this clearly is not the solution we desire. We therefore appeal to Box constrained optimization \citep{Box_1965} so that an upper bound can be enforced, thus preventing the optimization routine from drifting into the territory of extremely large branch lengths. Here, we choose to implement an upper bound equal to the largest pairwise JC69 distance, which will always exceed each individual branch length.

\par We also explore added model complexity. The framework available through the use of the robust distances $e_{kl}$ readily allows for natural estimation of model parameters from more complex models of DNA evolution, simultaneously with branch length estimation. We demonstrate an example using the K80 model \citep{Kimura_1980}. While the JC69 model assumes that all substitutions occur at the same rate, the K80 model allows for varying rates of certain types of substitutions. Specifically, the rate of transitions ($T \leftrightarrow C$ or $A \leftrightarrow G$) is often higher than the rate of transversions ($T,C \leftrightarrow A,G$). As such, \citet{Kimura_1980} proposed the K80 model to parameterize the transition and transversion rates separately, as $\alpha$ and $\beta$ respectively. The substitution rate matrix for the K80 model is thus
\begin{eqnarray}
\label{eqn:k80Qmatls}
\boldsymbol{\Lambda}_{K80} = \begin{pmatrix}
-(\alpha+2\beta) & \alpha & \beta & \beta \\
\alpha & -(\alpha+2\beta) & \beta & \beta \\
\beta & \beta & -(\alpha+2\beta) & \alpha \\
\beta & \beta & \alpha & -(\alpha+2\beta)
\end{pmatrix},
\end{eqnarray} 
where the row and column labels are in the following order: $(A, G, C, T)$. 
\par Frequently, the K80 model is parameterized in terms of the transition/transversion ratio $\kappa = \alpha/\beta$, giving rise to the rate matrix
\begin{eqnarray}
\label{eqn:k802}
\boldsymbol{\Lambda}_{K80} = \begin{pmatrix}
-(\kappa + 2) & \kappa & 1 & 1 \\
\kappa & -(\kappa + 2) & 1 & 1 \\
1 & 1 & -(\kappa + 2) & \kappa \\
1 & 1 & \kappa & -(\kappa + 2)
\end{pmatrix}.
\end{eqnarray} 
For least squares phylogenetic inference with the original loss function,
distances can be adjusted for $\kappa$; thus, the form of (\ref{eqn:L1}) remains the same, using adjusted
distance estimates $\hat{d}_{kl}^{(\kappa)}$. For our new loss function under the K80 model, we appeal to the concept of
labelled distances mentioned previously \citep{Minin_2008, Obrien_2009}, and count transitions and transversions separately. First, for our computations, we scale the K80 rate matrix by a factor of $1/(\kappa+2)$ so that the overall rate of leaving any given state is equal to 1, thereby setting time equal to the expected number of substitutions per site, so that our estimated robust distances can be interpreted directly as branch lengths in terms of time, as in our parameterization of the JC69 model described above. Thus, our K80 rate matrix is:
\begin{eqnarray}
\label{eqn:k803}
\boldsymbol{\Lambda}_{K80} = \begin{pmatrix}
-1 & \kappa/(\kappa + 2) & 1/(\kappa + 2) & 1/(\kappa + 2) \\
\kappa/(\kappa + 2) & -1 & 1/(\kappa + 2) & 1/(\kappa + 2) \\
1/(\kappa + 2) & 1/(\kappa + 2) & -1 & \kappa/(\kappa + 2) \\
1/(\kappa + 2) & 1/(\kappa + 2) & \kappa/(\kappa + 2) & -1
\end{pmatrix}.
\end{eqnarray} 
Then, following \cite{Obrien_2009}, we define labelled transition and transversion distances as:
\begin{eqnarray}
d_{(ts)} &=& E(N_t^{(ts)}) = t \cdot \sum_{i=1}^M \pi_i \sum_{j \neq i}^M \lambda_{ij} 1_{\{(i, j) \in (ts)\}} = t \cdot \frac{\kappa}{\kappa + 2}, \\
d_{(tv)} &=& E(N_t^{(tv)}) = t \cdot \sum_{i=1}^M \pi_i \sum_{j \neq i}^M \lambda_{ij} 1_{\{(i, j) \in (tv)\}} = t \cdot \frac{2}{\kappa +2},
\end{eqnarray}
where $(i, j) \in (ts)$ indicates that $i \rightarrow j$ is a transition, and $(i, j) \in (tv)$ indicates that $i \rightarrow j$ is a transversion. Then the labelled robust distances are:
\begin{eqnarray}
\label{eqn:robustd2}
e_{kl}^{(ts)}(\boldsymbol{\tau, b, \theta}) &=& \frac{1}{L} \sum_{s=1}^L \mathbb{E}(N_{t_{kl}(\boldsymbol{\tau},\boldsymbol{b})}^{(ts)}|y_{ks}, y_{ls}), \\
e_{kl}^{(tv)}(\boldsymbol{\tau, b, \theta}) &=& \frac{1}{L} \sum_{s=1}^L \mathbb{E}(N_{t_{kl}(\boldsymbol{\tau},\boldsymbol{b})}^{(tv)}|y_{ks}, y_{ls}),
\end{eqnarray}
which are calculated again following details laid out by \citet{Ball_2005}, \citet{Minin_2008} and \citet{Obrien_2009}. Finally, we define labelled transition and transversion tree distances as 
\begin{eqnarray}
t_{kl}^{(ts)}(\boldsymbol{\tau},\boldsymbol{b}) &=&  
t_{kl}(\boldsymbol{\tau},\boldsymbol{b}) \cdot \frac{\kappa}{\kappa+2}, \\
t_{kl}^{(tv)}(\boldsymbol{\tau},\boldsymbol{b}) &=&  
t_{kl}(\boldsymbol{\tau},\boldsymbol{b}) \cdot \frac{2}{\kappa+2},
\end{eqnarray}
so that each labelled tree distance is essentially a weighted proportion of the total tree distance due to transitions and transversions, respectively. Our K80 loss function is then
\begin{eqnarray}
\label{eqn:k80loss}
\mathcal{L}_3 = \sum_{k=1}^n \sum_{l=1}^n \left[e_{kl}^{(ts)} - t_{kl}^{(ts)}(\boldsymbol{\tau, b})\right]^2 + \left[e_{kl}^{(tv)} - t_{kl}^{(tv)}(\boldsymbol{\tau, b})\right]^2,
\end{eqnarray}
which we optimize over branch lengths and the transition/transversion ratio.

\paragraph{Implementation}
We have implemented our new least squares estimation method in the R package \texttt{phyloLSnoDist}, available at
\url{https://github.com/vnminin/phyloLSnoDist}. 
The package vignettes allow an interested reader to reproduce all numerical results reported in this paper.

\section{Results}

\subsection{Simulations}
\par To assess performance of phylogenetic inference with our proposed least squares criterion, we simulate nucleotide sequences using the R package phangorn \citep{Schliep_2011}. A variety of scenarios are considered,  in which we vary the number of taxa, the topology, and branch lengths. Throughout this work, we will use the following nomenclature to refer to particular scenarios: 
\begin{itemize}
\item Balanced tree (B) - all branch lengths are similar to each other
\item Unbalanced Long External 1 (ULE1) - one long external branch
\item Unbalanced Long External 2 (ULE2) - two long external branches
\item Unbalanced Long Internal 1 (ULI1) - one long internal branch 
\item Unbalanced Long Internal 2 (ULI2) - two long internal branches
\item Unbalanced Short Internal (USI1) - one short internal branch
\end{itemize}

\par We start by considering estimation of only the branch lengths. That is, we simulate DNA sequence alignments along a particular topology $\boldsymbol{\tau}$ and branch length set $\boldsymbol{b}$, and provide $\boldsymbol{\tau}$ to the estimation procedure as known and fixed. The topologies used for simulation range in size from five to seven taxa, with various branch length patterns as described above. In particular, it has been noted that the performance of phylogenetic inference with distance-based methods is inversely related to the diameter of the tree \citep{Atteson_1999, Lacey_2006, Roch_2010}, where diameter refers to the maximum distance between any two tips. Thus we include scenarios in which the true phylogeny has a large diameter (e.g. ULE1, ULE2, ULI1, ULI2). In each case, we use the JC69 model of DNA evolution.

\par
\begin{figure}
\begin{center}
\includegraphics[width=\textwidth]{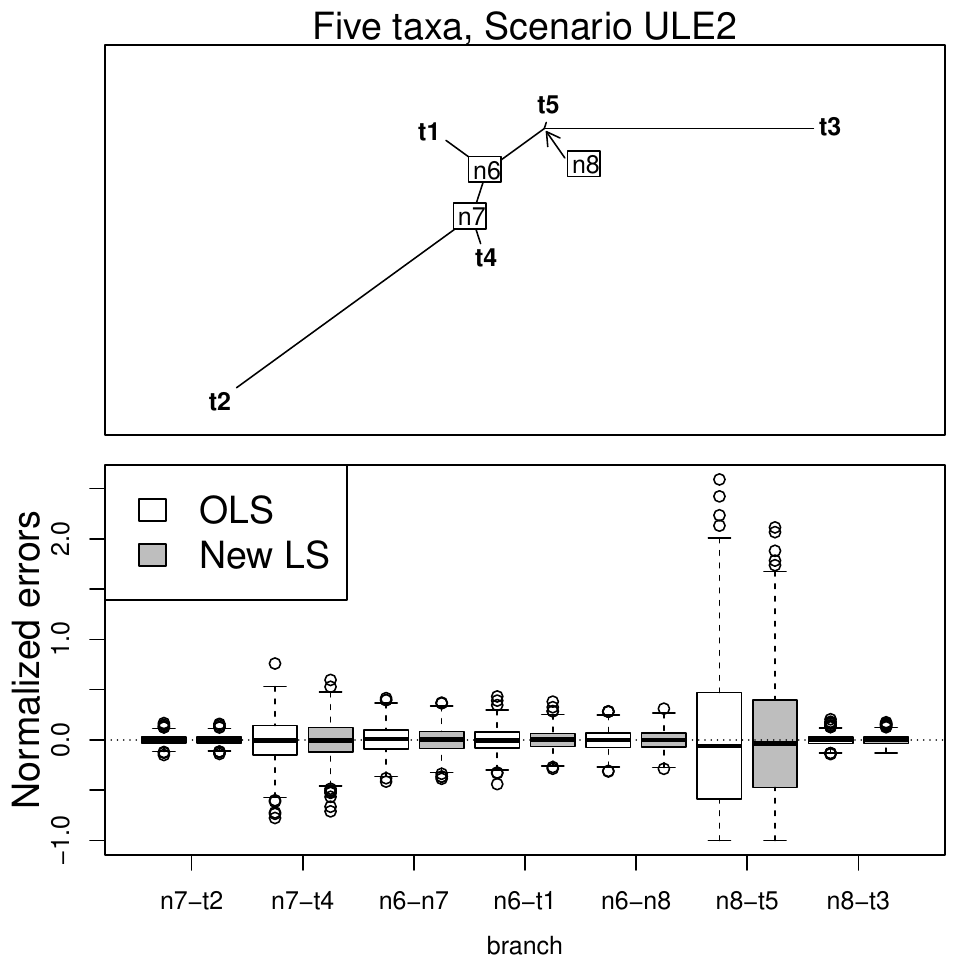}
\end{center}
\caption{\textbf{Simulation results: five taxa ULE2}. The tree used for simulations under this scenario is shown
in the top panel. For each method, normalized errors are calculated as $(\hat{b}_i - b_i^{\text{true}})/b_i^{\text{true}}$ for each
branch $b_i$. We compare boxplots of the normalized errors from 1000 iterations using ordinary least squares and our new approach, corresponding to optimization of the $\mathcal{L}_1$ and $\mathcal{L}_2$ loss functions,respectively.}
\label{fig:5-9}
\end{figure}

 First, we examine branch length estimation over the correct topology, through simulation of DNA sequence alignments along topologies of five, six and seven taxa, using the JC69 model and sequence lengths of 1000 nucleotides. Results from 1000 iterations with the five taxa, ULE2 scenario are shown in Figure \ref{fig:5-9} for all branches of the unrooted tree. We show boxplots of the normalized errors from each iteration, which are calculated as
\begin{eqnarray}
\text{Normalized error} = \frac{\hat{b}_i - b_i^{\text{true}}}{b_i^{\text{true}}},
\end{eqnarray}
for each branch $b_i$ and each method. 

\label{text:simLS}

\par
\begin{figure}
\begin{center}
\includegraphics[width=\textwidth]{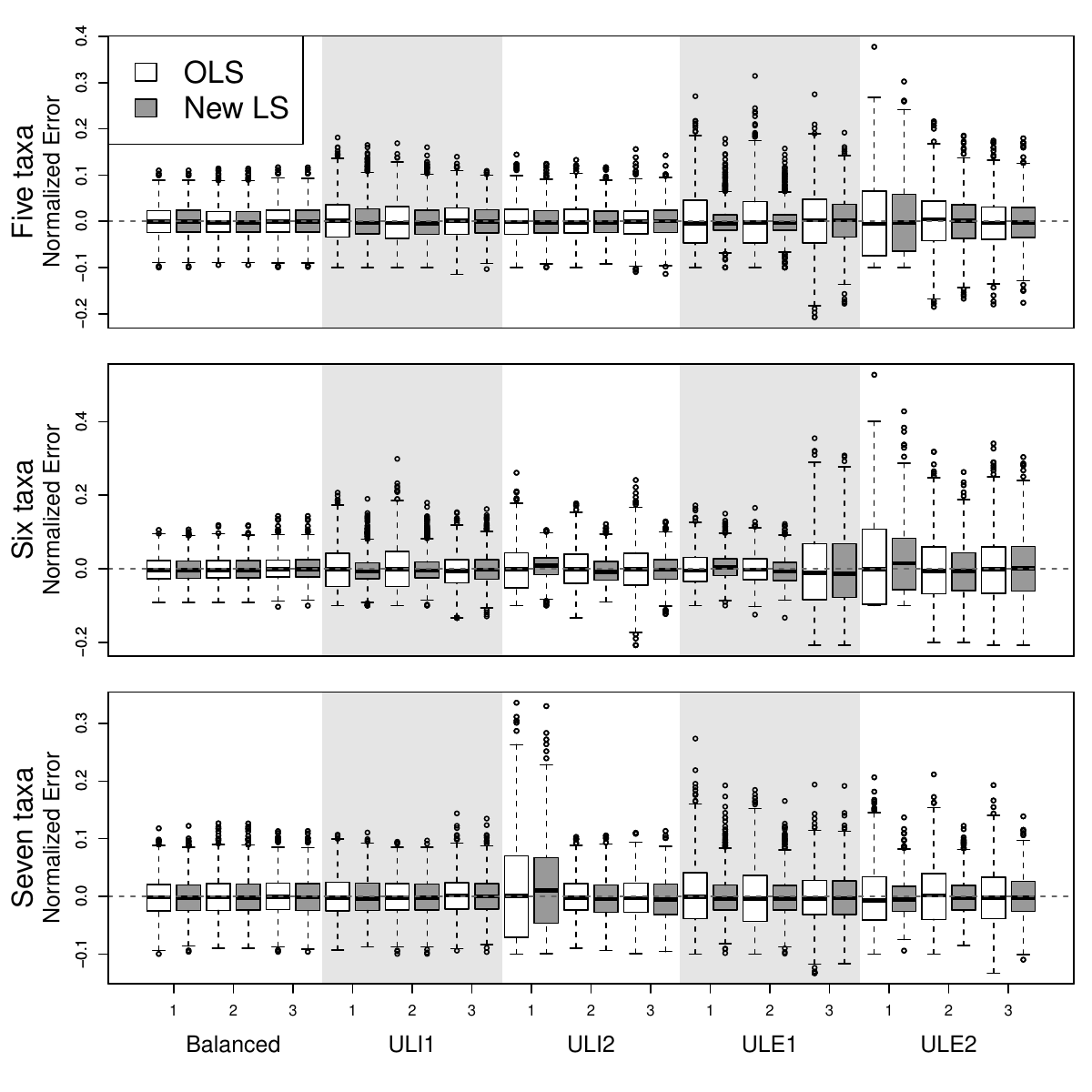}
\end{center}
\caption{\textbf{Simulation results: fixed topology}. We report boxplots of the normalized error for the shortest three branches (according to the true
tree) for each scenario, utilizing sequence lengths of 1000, and 1000 replicates for each scenario. Normalized Error is equal to $(\hat{b}_i - b_i^{\text{true}})/b_i^{\text{true}}$ for each branch $b_i$. Comparisons are made between ordinary least squares and our new approach, corresponding to optimization of the $\mathcal{L}_1$ and $\mathcal{L}_2$ loss functions, respectively.}
\label{fig:LS1}
\end{figure}

 \label{newLS1} For each scenario across the five, six and seven taxa, we show the normalized error for branch length estimates on the three shortest
branches of the true tree, in Figure \ref{fig:LS1} (again utilizing 1000 iterations and sequence lengths of 1000 nucleotides for each scenario). We observe that there is sometimes an improvement with
respect to the bias, but always an improvement with respect to the variability, which is shown in most of the branch
comparisons by the narrower interquartile range resulting from optimization of our new $\mathcal{L}_2$ loss function compared to ordinary least squares under the $\mathcal{L}_1$ loss function.

\par Next, we examine simultaneous topology and branch length estimation. Again, we simulate DNA sequence alignments along a particular topology $\boldsymbol{\tau}$ and branch length set $\boldsymbol{b}$ according to the JC69 model of DNA evolution, but now we do not assume that $\boldsymbol{\tau}$ is known. It has been noted that trees with short internal branch lengths are more difficult to infer correctly \citep{Martyn_2012}; thus, we use the USI1 scenario and compare how many times the correct topology is obtained by each method over a variety of sequence lengths. With four, five and six taxa trees, we simulate DNA sequence alignment data under the JC69 model, with a variety of sequence alignment lengths. We also include maximum likelihood estimation as a state-of-the-art benchmark. In the four and five taxa scenarios, an exhaustive search across all possible topologies is performed; in the six taxa scenario, a nearest-neighbor interchange (NNI) search \citep{Robinson_1971, Moore_1973} starting from the Neighbor-Joining tree \citep{Saitou_1987} is performed. Then, out of 100 replications, we count how many times each estimator obtained the correct topology in order to obtain the percentage of correct topology estimates.

\begin{figure}
\begin{center}
\includegraphics[width=\textwidth]{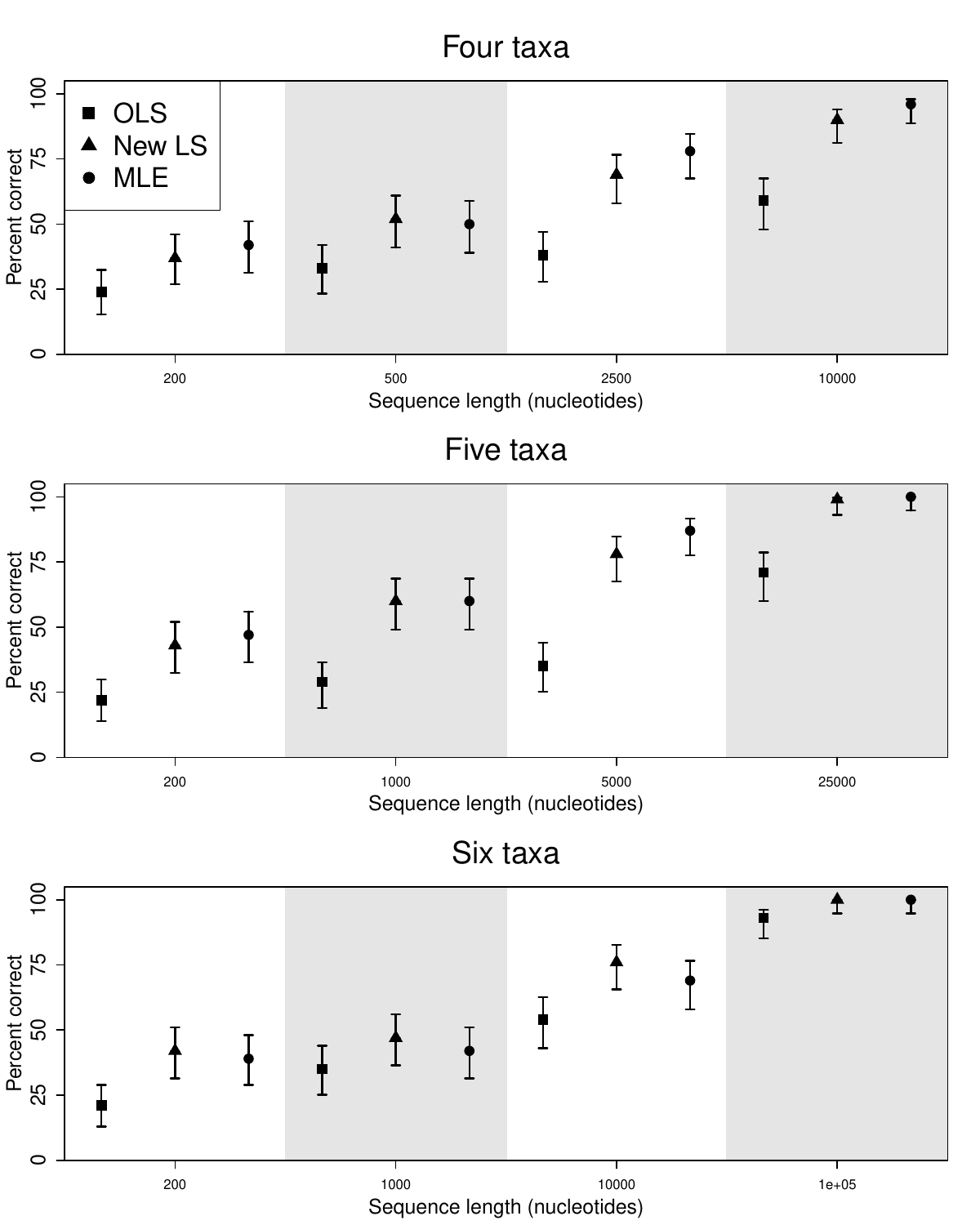}
\end{center}
\caption{\textbf{Simulation results: topology estimation}. Using the USI1 scenario, we estimate percentages of correct topology estimation across simulated sequence alignments of various lengths, with ordinary least squares (corresponding to optimization of the $\mathcal{L}_1$ loss function), our new least squares approach (corresponding to optimization of the $\mathcal{L}_2$ loss function), and maximum likelihoood estimation. Error bars represent LCO $95\%$ confidence intervals. For each scenario, we simulated 100 datasets.}
\label{fig:topol}
\end{figure}

\par Results are shown in Figure \ref{fig:topol}, with 95\% confidence intervals constructed according to the Length/Coverage optimal (LCO) method as it has been shown to have desirable statistical properties over the usual Wald interval for proportions, particularly as proportion point estimates approach 0 or 1 \citep{Schilling2014}. We also note that the set of sequence lengths used across the different number of taxa are not the same; this is because once the sequence length is sufficiently large, all three methods will result in nearly 100\% accuracy, which therefore will not illustrate any differences in the performance of the three methods. Moreover, the length at which this occurs is different for each number of taxa; that is, with four taxa trees, nearly 100\% accuracy will be obtained by all three methods at shorter sequence lengths than for five or six taxa trees. Thus, for each number of taxa, we choose a set of sequence lengths that best illustrates the gains that are possible using our $\mathcal{L}_2$ loss function as compared to ordinary least squares phylogenetic inference with the $\mathcal{L}_1$ loss function. 

\par In every scenario, our point estimate for the percentage of topologies correctly identified is higher using the $\mathcal{L}_2$ loss
function than the $\mathcal{L}_1$ loss function. However, we note that the $95\%$ confidence intervals within many scenarios overlap. Still, we note that the direction of the effect is the same all twelve of the scenarios, with seven of them being different enough to have non-overlapping confidence intervals between the estimation resulting from the $\mathcal{L}_1$ and $\mathcal{L}_2$ loss functions, and in three of the six-taxa scenarios, the $\mathcal{L}_2$ loss function is actually observed to perform better than maximum likelihood estimation, though again with overlapping confidence intervals. 

\par
\begin{figure}
\begin{center}
\includegraphics[width=\textwidth]{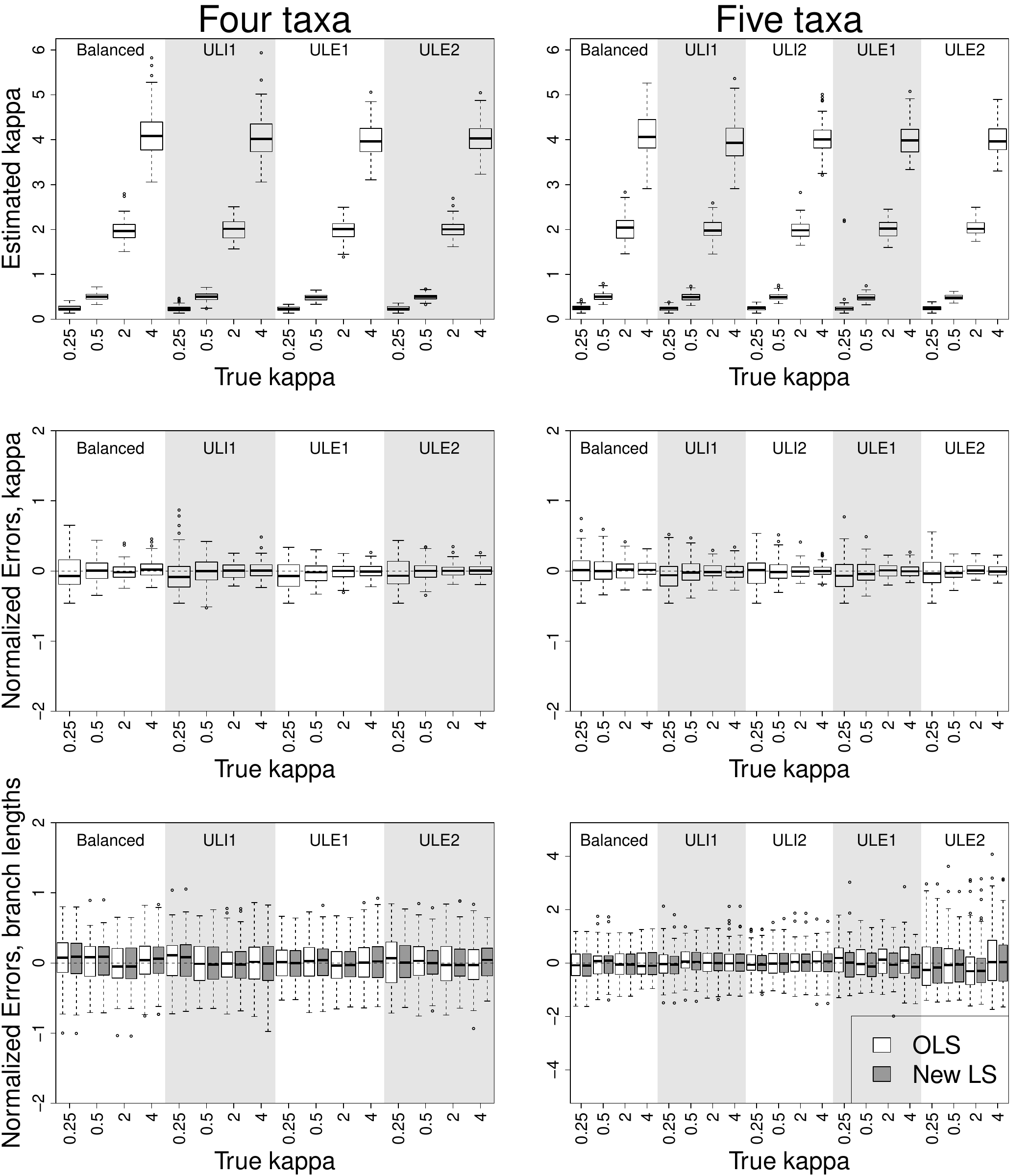}
\end{center}
\caption{\textbf{Simulation results: K80}. Boxplots from 100 replicates. Top panels show boxplots of $\kappa$ estimates from each scenario over true $\kappa$ values of 0.25, 0.5, 2 and 4. Middle panels show boxplots of normalized errors of $\kappa$ estimates, $(\hat{\kappa} - \kappa^{\text{true}}) / \kappa^{\text{true}}$. Bottom panels show normalized errors of the sum of all branch lengths in each scenario, comparing ordinary least squares to our new approach, corresponding to optimization of the $\mathcal{L}_1$ and $\mathcal{L}_2$ loss functions, respectively.}
\label{fig:LS-kappa}
\end{figure}

 Moving away from the JC69 model but returning to fixed topologies, we explore branch length optimization under the K80 model. Using transition/transversion ratios of  0.25, 0.5, 2 and 4, we simulate DNA sequence alignment data with four and five taxa, under the B, ULI1, ULE1 and ULE2 scenarios,
with sequence lengths of 1000 nucleotides, and 100 replications. Under five taxa, we also include the ULI2 scenario, but recalling that the ULI2 scenario is one with two long internal branches, this cannot be included for four taxa since unrooted four taxa trees do not have two internal branches. Summaries of the $\kappa$ estimates from each scenario are shown in the top panels of Figure \ref{fig:LS-kappa}, with normalized errors in the middle panels. The bottom panels show normalized errors for the total sum of all branches in the phylogeny.
We obtain $\kappa$ estimates that are well centered around the true values in each scenario, with normalized errors centered around zero. For the normalized errors of the total sum of branch lengths, the most consistent gains are obtained in the ULE2 scenario, with $\mathcal{L}_2$ boxplots exhibiting medians that are closer to 0 and smaller IQR than those of $\mathcal{L}_1$ in each case. 

\subsection{Data Analysis: Belgian HIV Transmission Chain}
\par Previously, \citet{Lemey_2005} and \citet{Vrancken_2014} studied the evolution of genes in the HIV virus in a Belgian sample of 11 HIV-positive patients, where the transmission chain was known from patient interviews and clinical information. One of the genes studied was the \textit{env} gene, which encodes for the spike protein and is thus responsible for the virus' ability to infiltrate host cells \citep{Coffin_1997}. A phylogenetic reconstruction using the \textit{env} gene matched the known transmission chain, thus demonstrating that phylogenetic reconstruction could be useful for uncovering an HIV transmission chain when it is not already known. 

\par Here, we revisit a subsample of \textit{env} genes from the dataset utilized by \citet{Vrancken_2014}. Specifically, we investigate the individuals identified as Patients A, B, F, H and I. Previously, contact tracing indicated that Patient A transmitted to Patient F, and Patient B transmitted to both Patients H and I \citep{Lemey_2005}. Prior to these transmission events, Patients A and B were both already HIV-positive and it was determined that one of them infected the other, but the time and direction of this transmission could not be established. For each individual, we chose to use the earliest available time point. Additionally, multiple clones for each time point were available for all individuals; we chose the ones that had both the fewest gap sites and no ambiguous sites. Our final dataset consisted of the sequences identified as A96cl7, B90cl22, F02cl15, H96cl02, and I99cl29 according to \citet{Vrancken_2014}. Any site with at least one sequence containing a gap was removed (resulting in 27 sites removed, out of 1557 total).

\begin{figure}
\begin{center}
\includegraphics[width=\textwidth]{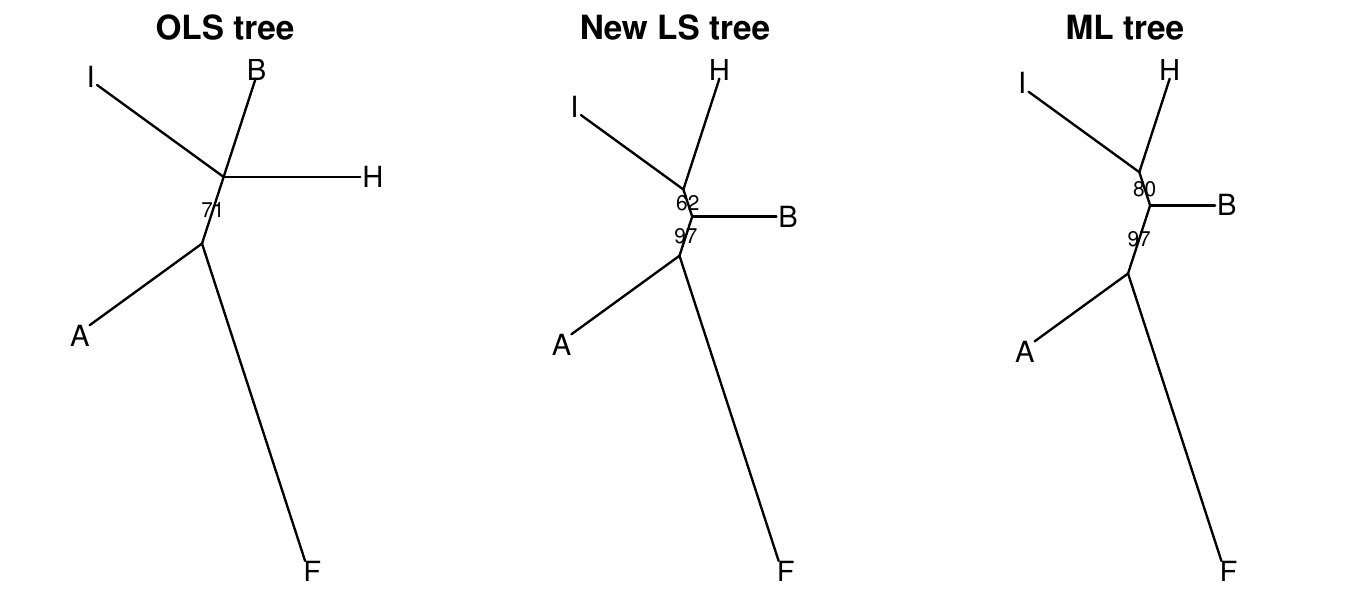}
\end{center}
\caption{\textbf{Phylogenies of the \textit{env} gene from Belgian HIV patients}. Phylogenies were inferred using ordinary least squares (by optimization of the $\mathcal{L}_1$ loss function), our new least squares approach (by optimization of the $\mathcal{L}_2$ loss function), and maximum likelihood. Numbers on internal branches represent bootstrap support from 100 bootstrap samples. In the OLS tree, we omit bootstrap support on the trifurcation.}
\label{fig:env}
\end{figure}

\par Using these samples, we perform phylogenetic reconstruction using the $\mathcal{L}_1$ loss function, our new $\mathcal{L}_2$ loss function, and maximum likelihood estimation. The inferred phylogenies are shown in Figure \ref{fig:env}, with estimated branch lengths shown in Table \ref{tab:env}. Bootstrap support is shown on each split in Figure \ref{fig:env}, and 95\% bootstrap confidence intervals are shown with each estimated branch length in Table \ref{tab:env}, from 100 bootstrap samples.

\par We observe that the $\mathcal{L}_2$ and ML trees have the same topology, and one that is consistent with the known transmission history, whereas the OLS tree contains a trifurcation leading to individuals I, B and H. Additionally, the ML tree has higher bootstrap support for the node leading to individuals I and H than the $\mathcal{L}_2$ tree. These differences are consistent with what is to be 
expected, with ML demonstrating the best performance, followed by the $\mathcal{L}_2$ tree and finally the $\mathcal{L}_1$ tree.

\begin{table}[ht]
\centering
\begin{tabular}{rccc}
  \hline
 & OLS tree & New LS tree & ML tree \\ 
  \hline
branch to I & 0.026 (0.017, 0.031) & 0.022 (0.015, 0.028) & 0.025 (0.018, 0.034) \\ 
  branch to H & 0.017 (0.0098, 0.021) & 0.02 (0.013, 0.028) & 0.018 (0.01, 0.026) \\ 
  interior branch 1 & 0 (-3.5e-18, 0) & 0.0049 (3.7e-06, 0.0071) & 0.0065 (0.002, 0.011) \\ 
  branch to B & 0.022 (0.015, 0.029) & 0.014 (0.0092, 0.02) & 0.012 (0.0068, 0.019) \\ 
  interior branch 2 & 0.011 (0.005, 0.02) & 0.0071 (0.0032, 0.017) & 0.013 (0.0071, 0.022) \\ 
  branch to A & 0.023 (0.017, 0.028) & 0.023 (0.018, 0.029) & 0.021 (0.016, 0.029) \\ 
  branch to F & 0.054 (0.044, 0.067) & 0.055 (0.044, 0.068) & 0.056 (0.046, 0.069) \\ 
   \hline
\end{tabular}
\caption{\textbf{Estimated branch lengths and 95\% bootstrap confidence intervals}. Estimated branch lengths correspond to their graphical representation in Figure \ref{fig:env}. Bootstrap confidence intervals, shown in parentheses, are obtained from 100 bootstrap samples.} 
\label{tab:env}
\end{table}

\section{Discussion}
\par Here, we propose a new least squares phylogenetic inference method. We find that for branch length estimation under the simple JC69 model of DNA evolution, modest gains can be achieved primarily with respect to variance, as compared to the regular least squares estimator. Much more substantial gains are observed for topology estimation, in which our approach results in a greater number of times that the correct topology was identified than that of regular least squares, and even is seen to surpass that of maximum likelihood estimation in a few cases, as seen in Figure \ref{fig:topol}.

\par One drawback to our use of the $\mathcal{L}_2$ loss function is that it is slower than the original least squares estimator using the $\mathcal{L}_1$ loss function, as traditional least squares phylogenetic inference can be accomplished with simple matrix operations as an ordinary least squares problem, solutions are nearly instantaneous, whereas the $\mathcal{L}_2$ loss function requires numeric optimization which can take more time. A potential answer could be the development of a more appropriate optimization routine to minimize the $\mathcal{L}_2$ loss function. Currently, we use the {\tt{nlminb}} routine, and while it was the best performer among all those that we tested, we note that it was not specifically designed for this optimization problem. For example, \citet{Felsenstein_1997} proposed and developed an optimization routine specifically for least squares phylogenetic inference. A similar development of an optimization routine specific to our loss function would likely result in substantial gains both in terms of speed and convergence.

\par Another issue arises from the fact that we must use box-constrained optimization for our new least squares estimator using $\mathcal{L}_2$. This is because, as mentioned previously, $\mathcal{L}_2 \rightarrow 0$ as each $t_{kl}(\boldsymbol{\tau},\boldsymbol{b}) \rightarrow \infty$. In our simulation studies, the usage of upper constraints has successfully produced well-behaved estimates of branch lengths, as we showed. However, it is fair to ask how one might know what to choose as upper constraint. Here, we choose an upper constraint equal to the largest pairwise JC69 distance from sequence alignment, which is generally well above the length of any individual branch length in a given tree. While this worked well here, it remains to be seen whether a more optimal solution exists.


\par It can be of interest to base phylogenetic inference on synonymous substitutions if we want to avoid the footprint of selective pressures, as demonstrated in \citep{Chi_2015}. Currently, least squares inference is the only choice when one wishes to distinguish between synonymous and nonsynonymous substitutions; the incorporation of labeled substitutions has not yet been demonstrated in the likelihood setting. Within our new framework proposed here, it is a natural extension to estimate a phylogeny based on synonymous substitutions. This may be desired if the aim is to infer a phylogeny on a group of taxa that is known to be under convergent selective pressures, in order to represent the true evolutionary history as opposed to one that is influenced by selection.

\par We have shown a proof of principle in this study, that least squares phylogenetic inference can be improved using our new loss function, in terms of branch length and topology estimation. Our gains are due to the fact that our robust distances are functions of both the sequence data and candidate tree, and the sequence data and candidate tree are considered simultaneously in the optimization of our loss function. It is also important to note that our approach allows for simultaneous estimation of branch lengths, topology, and substitution parameters under the framework of robust distances. As such, we demonstrate the added modeling flexibility under this framework, using the K80 model as an example. 
Moreover, we do not have to condition on pairwise alignments when forming the loss function \eqref{eqn:newloss}. 
\citet{minin2008fast} demonstrated how to compute the expected number of labeled substitutions along branches of a phylogenetic tree conditional on a multiple sequence alignment, so that our loss function could be extended to incorporate these results.

\section{Acknowledgements}
This work used the Augie High-Performance Computing cluster, funded by award NSF 2018933, at Villanova University.

\bibliography{pc}{}

\clearpage
\renewcommand{\thesection}{S}  
\renewcommand{\thetable}{S\arabic{table}} 
\setcounter{table}{0}
\section{Supplementary Material}
\subsection{Numerical Optimization Routines}

\par In order to implement our desired optimization, we turn to currently available packages in {\tt{R}}. We compare
the performance of seven routines which are capable of iterated box-constrained optimization over several variables 
(Table \ref{tab:optimizers}), both in their ability to actually minimize the loss function, and in their speed. We simulated DNA sequence alignments on seven taxa trees, and ran each optimization routine with our loss function $\mathcal{L}_2$.

\par Although each routine has convergence criteria that it aims to satisfy in order to report whether the minimum has been
obtained or not, any given two routines did not necessarily agree on the returned loss function value, even when both had 
independently determined that it has achieved convergence. Thus, in comparing these seven routines, we deem that convergence has been
achieved by the one which returned the smallest value of the loss function, and record this under the ``$\#$ best'' column
in Table \ref{tab:optimizers}. The average time per iteration
is shown under ``Time (s).'' 

\begin{table}[h]
\begin{center}
\caption{\textbf{Performance of box-constrained optimizers}. ``\# best'' indicates the count of how many
times each routine found the smallest loss function value. ``Time (s)'' is the average time of each routine, in seconds,
across all 30 iterations performed.}
\vspace{2mm}
\label{tab:optimizers}
\begin{tabular}{llccl}
  \hline
Algorithm & Package & \# best & Time (s) & Reference \\ 
  \hline
{\tt{bobyqa}} & \textbf{minqa} & 0 & 137.5 & \citep{Powell_2009} \\ 
  {\tt{L-BFGS-B}} & \textbf{stats} & 0 & 47.6 & \citep{Byrd_1995} \\ 
  {\tt{nlminb}} & \textbf{stats} & 26 & 25.5 & \citep{Gay_1990, Fox_1997} \\ 
  {\tt{nmk}} & \textbf{dfoptim} & 0 & 22.0 & \citep{Kelley_1999} \\ 
  {\tt{Rcgmin}} & \textbf{Rcgmin} & 0 & 55.4 & \citep{Nash_1979} \\ 
  {\tt{Rvmmin}} & \textbf{Rvmmin} & 4 & 33.3 & \citep{Nash_1979} \\ 
  {\tt{spg}} & \textbf{BB} & 0 & 137.3 & \citep{Birgin_2000} \\ 
   \hline
\end{tabular}
\end{center}
\end{table}

\par Out of 30 iterations performed, {\tt{nlminb}} found the smallest value 26 times, and 
{\tt{Rvmmin}} found the smallest value 4 times. We note that although we did not observe any 
ties in the minimized loss function value, in most cases 3-4 of the routines would agree within 10 decimal places. This
in fact was the case with {\tt{nlminb}} and {\tt{Rvmmin}} in the 4 times when {\tt{Rvmmin}} found the smallest value. On the other
hand, {\tt{nmk}} (the fastest routine, on average) often returned values far above the smallest value, sometimes as much as
50 times greater. Since {\tt{nlminb}} is the next fastest at an average of 25.5 seconds per iteration, and has excellent
performance as just previously noted, we choose this routine to optimize our loss function $\mathcal{L}_2$.

\subsection{Analytic expressions for $\mathcal{L}_2$ loss function under JC69 Model}
Recalling that 
\begin{eqnarray}
\mathcal{L}_2(\boldsymbol{Y, b, \tau, \theta}) &=& \sum_{k > l}\bigg[ e_{kl}(\boldsymbol{\tau, b, \theta}) - t_{kl}(\boldsymbol{\tau, b}) \bigg]^2, 
\end{eqnarray}
where
\begin{eqnarray}
e_{kl}(\boldsymbol{\tau, b, \theta}) = \frac{1}{L} \sum_{s=1}^L \mathbb{E}(N_{t_{kl}(\boldsymbol{\tau},\boldsymbol{b})} | y_{ks}, y_{ls}),
\end{eqnarray}
we now specify this for the JC69 model. From the rate matrix also shown previously in Equation (\ref{jc69rate}),
\begin{align}
\boldsymbol{\Lambda}_{JC} &= \frac{1}{4} \begin{pmatrix}
-3 \alpha & \alpha & \alpha & \alpha \\
\alpha & -3 \alpha & \alpha & \alpha \\
\alpha & \alpha & -3 \alpha & \alpha \\
\alpha & \alpha & \alpha & -3 \alpha \\
\end{pmatrix},
\end{align}
we obtain transition probabilities:
\begin{align}
\label{jc69prob2}
P(X_t = j | X_0 = i) \equiv p_{ij}(\alpha, t) = 
\left\{
\begin{array}{ll}
\frac{1}{4}\bigg(1 + 3e^{-\alpha t}\bigg) &\text{ for } i = j \\
\frac{1}{4}\bigg(1 - e^{-\alpha t}\bigg) &\text{ for } i \neq j,
\end{array}
\right.
\end{align}
which matches the transition probability matrix shown in Equation (\ref{jc69prob}). Now, \cite{Ball_2005} and \cite{Minin_2008} showed that
\begin{align}
\mathbb{E}(N_t \cdot 1_{\{X_t = j\}} | X_0 = i) = t \cdot [\boldsymbol{\Lambda} - \text{diag}(\boldsymbol{\Lambda})] e^{\boldsymbol{\Lambda} t},
\end{align}
which, for the JC69 model, is equal to:
\begin{align}
t \cdot \begin{pmatrix}
0 & \alpha & \alpha & \alpha \\
\alpha & 0 & \alpha & \alpha \\
\alpha & \alpha & 0 & \alpha \\
\alpha & \alpha & \alpha & 0 \\
\end{pmatrix} \times 
\frac{1}{4} \begin{pmatrix}
1+3 e^{- \alpha t} & 1-e^{-\alpha t} & 1-e^{-\alpha t} & 1-e^{-\alpha t} \\
1-e^{-\alpha t} & 1+3 e^{- \alpha t} & 1-e^{-\alpha t} & 1-e^{-\alpha t} \\
1-e^{-\alpha t} & 1-e^{-\alpha t} & 1+3 e^{- \alpha t} & 1-e^{-\alpha t} \\
1-e^{-\alpha t} & 1-e^{-\alpha t} & 1-e^{-\alpha t} & 1+3 e^{- \alpha t}
\end{pmatrix},
\end{align}
giving rise to:
\begin{align}
\mathbb{E}(N_t \cdot 1_{\{X_t = j\}} | X_0 = i) = 
\left\{ \begin{array}{ll}
t \cdot \frac{3 \alpha}{4}\bigg(1 - e^{-\alpha t}\bigg) &\text{for } i = j \\
t \cdot \frac{\alpha}{4} \bigg[(1 + 3e^{-\alpha t}) + 2(1-e^{-\alpha t})\bigg] &\text{for } i \neq j.
\end{array}
\right.
\end{align}
Then, with $p_{ij}(\alpha, t)$ as defined in Equation (\ref{jc69prob2}), by the definition of conditional expectation we have:
\begin{align}
\mathbb{E}(N_t|X_t=j, X_0=i) = \frac{\mathbb{E}(N_t \cdot 1_{\{X_t=j\}} | X_0=i)}{p_{ij}(\alpha, t)} \equiv m_{ij}(\alpha, t) = \left\{
\begin{array}{ll}
t \cdot \frac{3 \alpha}{4} \frac{p_{12}(\alpha, t)}{p_{11}(\alpha, t)} &\text{for } i = j \\
t \cdot \frac{\alpha}{4} \bigg[2 + \frac{p_{11}(\alpha, t)}{p_{12}(\alpha, t)}\bigg] &\text{for } i \neq j.
\end{array}
\right.
\end{align}
Now, for identifiability we set $\alpha=(4/3)$, so that
\begin{align}
E(N_t) = \sum_{i=1}^M \pi_i \lambda_{i} t = \sum_{i=1}^M \frac{1}{4} \bigg(\frac{3}{4} \alpha \bigg) t = 4\bigg(\frac{1}{4}\bigg) \bigg(\frac{3}{4} \cdot \frac{4}{3} \bigg) t = t,
\end{align}
meaning that time is measured in the expected number of substitutions per site. Then, 
\begin{align}
e_{kl}(\boldsymbol{\tau, b, \theta}) &= \frac{1}{L} \sum_{s=1}^L \mathbb{E}(N_{t_{kl}(\boldsymbol{\tau},\boldsymbol{b})} | y_{ks}, y_{ls}) \\
&{\overset{\text{JC69}}=} \frac{1}{L} \bigg[n^0_{kl} 
\times m_{11}\bigg(4/3, t_{kl}(\boldsymbol{\tau, b})\bigg) + 
n^1_{kl} \times m_{12}\bigg(4/3, t_{kl}(\boldsymbol{\tau, b})\bigg)
\bigg],
\end{align}
where 
\begin{align}
n_{kl}^0 &= \sum_{s=1}^L 1_{\{y_{ks} = y_{ls}\}}, \\
n_{kl}^1 &= L - n_{kl}^0,
\end{align}
that is, $n_{kl}^0$ is the number of invariant sites, and $n_{kl}^1$ is the number of variable sites, between taxa $k$ and $l$. Letting $d_{kl}^h = n_{kl}^1/L$, the robust distance $e_{kl}(\boldsymbol{\tau, b, \theta})$ under the JC69 model then becomes:
\begin{align}
e_{kl}(\boldsymbol{\tau, b, \theta}) &= (1-d_{kl}^h) \cdot m_{11}\bigg(4/3, t_{kl}(\boldsymbol{\tau, b})\bigg) + d_{kl}^h \cdot m_{12}\bigg(4/3, t_{kl}(\boldsymbol{\tau, b})\bigg),
\end{align}
and our loss function can be expressed as:
\begin{align}
\label{elementloss}
\mathcal{L}_2(\boldsymbol{Y, b, \tau, \theta}) &= \sum_{k > l}\bigg[ (1-d_{kl}^h) \cdot m_{11}\bigg(4/3, t_{kl}(\boldsymbol{\tau, b})\bigg) + d_{kl}^h \cdot m_{12}\bigg(4/3, t_{kl}(\boldsymbol{\tau, b})\bigg) - t_{kl}(\boldsymbol{\tau, b}) \bigg]^2.
\end{align}
In matrix form, we define lower-triangular matrices:
\begin{align}
\boldsymbol{D}^h &= \bigg\{d_{kl}^h \cdot 1_{\{k > l\}}\bigg\}, \\
\widetilde{\boldsymbol{D}}^h &= \bigg\{(1-d_{kl}^h) \cdot 1_{\{k > l\}}\bigg\}, \\
\boldsymbol{E} &= \bigg\{ e_{kl}(\boldsymbol{\tau, b, \theta}) \cdot 1_{\{k > l\}} \bigg\}, \\
\boldsymbol{M}_{11} &= \bigg\{m_{11}\bigg(4/3, t_{kl}(\boldsymbol{\tau, b})\bigg) \cdot 1_{\{k > l\}} \bigg\}, \\
\boldsymbol{M}_{12} &= \bigg\{m_{12}\bigg(4/3, t_{kl}(\boldsymbol{\tau, b})\bigg) \cdot 1_{\{k > l\}} \bigg\}, \\
\boldsymbol{T} &= \bigg\{t_{kl}(\boldsymbol{\tau, b}) \cdot 1_{\{k > l\}}\bigg\},
\end{align}
and
\begin{align}
\boldsymbol{H} &= \widetilde{\boldsymbol{D}}^h \odot \boldsymbol{M}_{11} + \boldsymbol{D}^h \odot \boldsymbol{M}_{12} - \boldsymbol{T},
\end{align}
where $\odot$ is the Hadamard element-wise product. Then, with $\boldsymbol{H} = \{h_{kl}\}$, our loss function can be expressed as:
\begin{align}
\mathcal{L}_2(\boldsymbol{Y, b, \tau, \theta}) &= \sum_{k>l} h^2_{kl}.
\end{align}
The matrices $\boldsymbol{D}^h$ and $\widetilde{\boldsymbol{D}}^h$ can be computed once and cached to be re-used during loss function optimization. 
\subsection{Proof of $\mathcal{L}_2 \rightarrow 0$ as each $t_{kl}(\boldsymbol{\tau},\boldsymbol{b}) \rightarrow \infty$, JC69 case}
Examining the element-wise form of our loss function shown in Equation (\ref{elementloss}), for any given pair of taxa $k$ and $l$, and replacing $t_{kl}(\boldsymbol{\tau, b})$ with $t$ due to the restriction that $\alpha=(4/3)$, we have:
\begin{align}
\mathcal{L}_2(\boldsymbol{Y, b, \tau, \theta}) &= \bigg[ (1-d_{kl}^h) \cdot m_{11}\bigg(4/3, t_{kl}(\boldsymbol{\tau, b})\bigg) + d_{kl}^h \cdot m_{12}\bigg(4/3, t_{kl}(\boldsymbol{\tau, b})\bigg) - t_{kl}(\boldsymbol{\tau, b}) \bigg]^2 \\
&= \bigg[(1-d_{kl}^h) \cdot t \cdot \frac{p_{12}(4/3, t)}{p_{11}(4/3, t)} + d_{kl}^h \cdot t \cdot \frac{1}{3} \left(2 + \frac{p_{11}(4/3, t)}{p_{12}(4/3, t)}\right) - t\bigg]^2 \\
&= \bigg[(1-d_{kl}^h) \cdot t \cdot \frac{1-e^{-(4/3) t}}{1+3e^{-(4/3) t}} + d_{kl}^h \cdot t \bigg(\frac{2}{3} + \frac{1+3e^{-(4/3) t}}{3(1-e^{-(4/3) t})} \bigg) - t(1-d_{kl}^h) - t(d_{kl}^h)  \bigg]^2 \\
&= \bigg[(1-d_{kl}^h) \bigg(t \cdot \frac{1-e^{-(4/3) t}}{1+3e^{-(4/3) t}} - t \bigg) + d_{kl}^h \bigg(t \cdot \frac{1+3e^{-(4/3) t}}{3(1-e^{-(4/3) t})} - \frac{t}{3} \bigg) \bigg]^2 \\
&=\bigg[(1-d_{kl}^h) \cdot t\bigg(\frac{1-e^{-(4/3)t}}{1+3e^{-(4/3)t}} - 1 \bigg) + d_{kl}^h \cdot t \bigg(\frac{1+3e^{-(4/3)t}}{3(1-e^{-(4/3)t})} - \frac{1}{3}\bigg) \bigg]^2 \\
&= \bigg[(1-d_{kl}^h) \bigg(\frac{-4te^{-(4/3)t}}{1+3e^{-(4/3)t}}\bigg) + d_{kl}^h \bigg(\frac{4te^{-(4/3)t}}{3(1-e^{-(4/3)t})}\bigg) \bigg]^2.
\end{align}
Noting that $te^{-(4/3)t} \rightarrow 0$ as $t \rightarrow \infty$, that each denominator in the above expression converges to a constant as $t \rightarrow \infty$, and that $d_{kl}^h$ is constant with respect to $t$, the proof is complete. 

Intuitively, this observation follows from the fact that, marginally, 
$\mathbb{E}\left[N_{t_{kl}(\boldsymbol{\tau, b})}\right]=t_{kl}(\boldsymbol{\tau, b})$ by definition; then, as $t_{kl}(\boldsymbol{\tau, b}) \rightarrow \infty$,
$N_{t_{kl}(\boldsymbol{\tau, b})}$ becomes independent of the observed states. This is because $N_{t_{kl}(\boldsymbol{\tau, b})}$ and the observed
states are both properties of the Markov chain path that rely on the amount of time that has passed. Specifically,
$N_{t_{kl}(\boldsymbol{\tau, b})}$ counts the number of jumps along the entire path, and $y_{ks}, y_{ls}$ can be considered as the start and end points, respectively. 
As the length of the path grows large, the count starts to rely less on where it began and ended.

\end{document}